\documentclass[epj]{svjour}
\usepackage{graphicx}
\usepackage{amsmath}
\usepackage{amsfonts}
\usepackage{amssymb}
\begin{document}
\title{Adaptation of fictional and online conversations to communication media}
\author{Christian M. Alis \and May T. Lim
}                     % Do not remove
%
%\offprints{}          % Insert a name or remove this line
%
\institute{National Institute of Physics, \\
University of the Philippines Diliman\\
1101 Quezon City, Philippines}
\date{Received: date / Revised version: date}
% The correct dates will be entered by Springer
%
\abstract{
Conversations allow the quick transfer of short bits of information and it is reasonable to expect that changes in communication medium affect how we converse. Using conversations in works of fiction and in an online social networking platform, we show that the utterance length of conversations is slowly shortening with time but adapts more strongly to the constraints of the communication medium. This indicates that the introduction of any new medium of communication can affect the way natural language evolves.
\PACS{
      {89.65.Ef}{Social organizations; anthropology}   \and
      {89.20.-a}{Interdisciplinary applications of physics}
     } % end of PACS codes
} %end of abstract
\maketitle
\section{Introduction}
\label{intro}
With an estimated vocabulary size of 20,000 to 40,000 base words~\cite{goulden_how_1990,nation_vocabulary_1997,browne_measuring_2007}, conversations quickly transfer short bits of information via two general means: the oral and the written form. Although the written vocabulary is often larger~\cite{hayes_vocabulary_1988}, the grammatically looser and more error-prone oral medium has the advantage of having access to nonverbal cues like gestures and intonations~\cite{hill_oral_2010} to aid communication. Aside from vocabulary size---word choices, unconsciously repeating words, and other idiosyncrasies~\cite{chafe_relation_1987} also affect the way we perceive conversations.

Conversation analysis typically looks into how turn taking patterns in institutional settings depart from those observed in informal conversations~\cite{wooffitt_conversation_2005}, or on the psychological or sociological aspects~\cite{sack_conversationmap_2000} of social structure. In this work, the length distribution of a single speaking turn, or utterance, was derived to determine if the medium affects the way we express ideas by using datasets that include a mix of real-world (online) and fictional (offline) conversations: online conversation in Twitter (\texttt{twitter.com}); conversations from 19th century novels and short stories; and subtitles from 20th century movies. 

Humans typically converse orally, thus the analysis  of conversations is usually performed by transcribing recorded audio conversations into text. In cases when this is not possible e.g., before the invention of recorded audio, one technique is to use written records of real and constructed conversations as were done in studies on the emergence of complementary clauses (Paul persuaded John \textit{to kiss Mary})~\cite{warner_complementation_1982}, the use of \textit{do} in negative declaratives (I \textit{do not} understand you)~\cite{warner_why_2005}, and the increasing prevalence of the modals \textit{gonna}, \textit{gotta} and \textit{wanna}~\cite{lorenz_emancipation_2012}. Written records of spoken speech are also included in corpora like \textit{A Corpus of English Dialogues 1560--1760}~\cite{kyto_guide_2006} and \textit{The Corpus of Historical American English: 400 million words, 1810-2009}~\cite{davies_corpus_2010}. However, only conversations (fictional dialogues) in novels, short stories, and movies were analyzed in this paper because utterances tend to be less narrative and directed to another person unlike in other genres like drama comedies or trial transcripts. Although it has been shown that styles vary across and even within authors~\cite{smith_stylistic_2002}, we assumed that conversations in their works are mostly independent of the author's style, i.e., a conversation in their works conveys how another person (character), and not how the author, speaks. Furthermore, errors due to transcribing are practically eliminated when using books and movies.

Twitter, as a form of computer-mediated communication, is different from oral or written media~\cite{crystal_language_2006}. While assumed to be happening in real-time, the purely written nature of a Twitter-based conversation differentiates it from the transcribed oral communication in books and movies. In addition, Twitter conversations have an explicit length limit---an utterance can only be up to 140 characters long. 

Putting a length constraint on the outset would show drastic changes. A case in point would be SMS messages. At its peak, \textit{textspeak} looked very much different from standard spelling---primarily due to the effort it takes to spell out words through a numerical keypad. Tweets, however, was largely spared from this phenomenon and usually have correct spelling. Among the three media analyzed in this study, Twitter is the only considered medium that is constrained. Conversations in books and movies are supposedly oral conversations that were written down in the form of a book or a subtitle so their written form should have no effect on them. 

We now argue that if conversations are independent of medium, then no significant difference should be observed among conversations in Twitter, books and movies. On the other hand, if differences in a medium is due to an explicit quirk in the medium e.g., an utterance length limit, then conversations in Twitter must be significantly different from conversations in books and movies, but the latter two should not be significantly different from each other. Finally, if conversations are indeed dependent on medium, then conversations in Twitter, movies, and books must be significantly different from each other.

\section{Orthographic sentence length and the Brown corpus}
\label{sec:ortho}
The study of sentence lengths in text dates back to the 1939 paper of Udny Yule~\cite{yule_sentence-_1939} where it was used to establish authorship. More recently, sentence length has been used to classify text genre by itself~\cite{spiliopoulou_quantitative_2006} or in combination with other text properties~\cite{copeck_automating_2000}. Yule's 1939 paper did not provide the sentence length distribution but several decades after its publication, the distribution was described as lognormal~\cite{williams_note_1940,williams_style_1970,wake_sentence-length_1957} which was later shown by Sichel~\cite{sichel_distribution_1974,grzybek_history_2006} to be flawed. More recently, Sigurd et al.~\cite{sigurd_word_2004} showed that sentence length distributions may be approximated by a gamma distribution.

In this work, we used the non-standard unit of number of characters (orthographic length), instead of the usual sentence length units of clauses or words, in measuring utterance lengths for ease of comparison with Twitter which has a maximum utterance length in terms of characters. Although the distribution of sentence lengths in terms of words and word lengths in terms of letters can be described by a gamma distribution~\cite{sigurd_word_2004}, there is no mathematical guarantee that the distribution of sentence lengths in terms of letters would also follow the same distribution in the general case of different shape and scale parameters of the sentence length (in words) and word length (in letters) distributions. If it can be shown that the sentence length (in letters) distribution can be approximated by a member of the same distribution family as the sentence length (in words), then the use of sentence length comparison using orthographic length is a valid approach.

The Brown corpus~\cite{kucera_computational_1967} consists of about one million words of edited English prose printed during 1961 in the United States~\cite{francis_brown_1979}. To verify if measuring sentence lengths in terms of characters may be approximated by a gamma distribution, the sentence length (in letters) distribution was simulated, as follows. The word length (in letters) and sentence length (in words) distributions of the tagged Brown corpus was first constructed using the natural language toolkit~\cite{bird_natural_2009} Python module. In constructing the distributions, only words that contain at least one letter were considered. A gamma distribution given by,
\begin{equation}
\label{eq:gammafit_sl}
\mathrm{Pr}(x) = \frac{x^{\alpha - 1} e^{x/s}}{ s^\alpha \Gamma(\alpha)},
\end{equation}
where $\alpha$ and $s$ are fitting parameters that describe the shape and ordinate scaling factor, respectively, were then fitted on each distribution using the maximum likelihood estimation~\cite{wasserman_all_2004} feature of the Scipy python module~\cite{jones_scipy:_2001}. For each trial, 100,000 sentences were generated following the fitted sentence length (in words) and word length (in letters) distributions. This process was repeated for a total of 100 trials resulting to 100 sentence length in letters histograms. The histograms were converted to a single probability distribution by using the median frequency for each sentence length.

\begin{figure}[!ht]
\centerline{\includegraphics[scale=0.5]{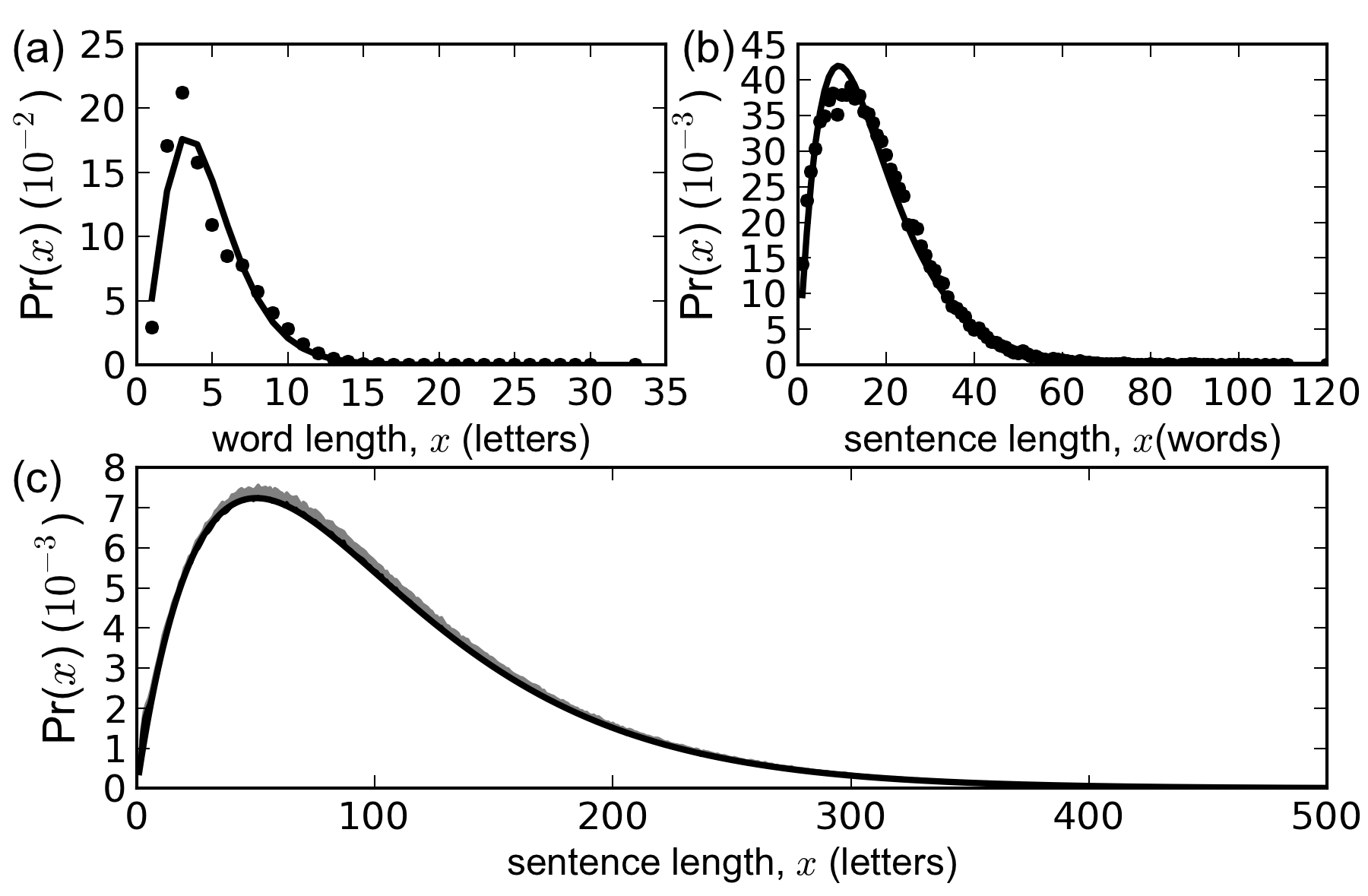}}
\caption{ (a) Word length in letters and (b) sentence length in words distributions of the Brown corpus superimposed with the maximum likelihood estimate of Eq.~\eqref{eq:gammafit_sl} (solid line). (c) Simulated sentence length (solid dots) in letters distribution using the fitted word length (in letters) and sentence length (in words) distributions of the Brown corpus superimposed with the least-squares fit (solid line) and values within one standard deviation (shaded)
}
\label{fig:wl-prob}
\end{figure}

Both the word length (WL, in letters) [Fig.~\ref{fig:wl-prob}(a)] and sentence length (SL, in words) [Fig.~\ref{fig:wl-prob}(c)] distributions of the Brown corpus follow a gamma distribution (WL: $\alpha$ = 3.43, $s$ = 1.39, $r^2$ = 0.948; SL: $\alpha$ = 2.09, $s$ = 8.44, $r^2$ = 0.989). The simulated sentence length in letters distribution [Fig.~\ref{fig:wl-prob}(b)] also follows a gamma distribution ($\alpha$ = 1.98, $s$ = 51.2) but has a much larger $s$ than the sentence length in words which is expected since letters is a smaller syntactic unit than words.

The sentence length distribution in letters thus belongs to the same family of distributions as when measured in words. Since utterance lengths are being compared empirically, the use of orthographic length as a unit of utterance length is therefore valid despite known idiosyncrasies~\cite{_english_2008} of the English language. Interestingly, the orthographic length was also used by Piantadosi et al.~\cite{piantadosi_word_2011} when they showed that word lengths are optimized for efficient communication because it is easier to measure while still being highly correlated with word length in terms of syllables~\cite{strauss_word_2006}.

\section{Datasets}

Four datasets were used for our analysis: utterances in fictional works in Project Gutenberg (\textsc{pg})(\texttt{gutenberg.org}), utterances in \textsc{pg} split into sentences (\textsc{pgs}), tweets from Twitter (\textsc{twitter}), and utterances in movie subtitles (\textsc{subs}) from \texttt{opensubtitles.org}.

\textsc{pg} was generated by extracting utterances---defined as text enclosed in double quotes---from the available works in Project Gutenberg of 50 authors whose selection was roughly based on availability (see Ref.~\cite{alis_pg_authorslist} for list of titles, and Ref.~\cite{alis_supplementary_2012} for author selection and text parsing details). The resulting dataset consists of about 2.3 million utterances, with zero-length utterances (0.01\% of original dataset) removed. The author with the most number of utterances (George Manville Fenn) has 238,640 utterances while the author with the least number of utterances (David Herbert Lawrence) has 1,170 utterances. The median number of utterances is 36,955 utterances per author. When split into sentences, \textsc{pg} is converted to \textsc{pgs} which has about 4.2 million utterances with a median number of utterances equal to 69,311 utterances per author.

Conversations in \textsc{twitter} were identified by looking for \textit{replies}, which are Twitter messages (or \textit{tweets}) directed to specific users. We used the convention that \textit{replies} begin with the \texttt{@username} of the receiver, e.g., \textit{@bob Hello! How are you?} to filter the tweets for our dataset.\footnote{The current Twitter API supports a method for explicitly classifying a tweet as a \textit{reply} but this was not yet widely available and followed when our data were gathered.} Though not in the original design, the use of \textit{replies} emerged as the leading method of addressing a particular person in Twitter~\cite{williams_how_2008}. The presence of an \texttt{@username} anywhere in the tweet makes that tweet a \textit{mention} \cite{_what_2012}.  Unlike \textit{mentions}, which appear in the timeline of a user following the sender, a \textit{reply} appears in said user's timeline only if he follows both sender and receiver of the \textit{reply} message. Thus, conversations are most likely restricted to \textit{replies} to avoid flooding the timeline of people not involved in the discussion. Though \textit{mentions} may carry conversations, we still excluded them from the dataset, as they are more likely non-conversational tweets.

It is possible that a \textit{reply} is not reciprocated, e.g., if it was meant to bring an item, such as a URL, to the attention of another user. This is still considered a conversation because it conveys a short bit of information directly targeted to a certain user. This is similar to someone telling another to ``watch out!" or ``be careful": a reply by the other person is not required.

Using the Twitter Streaming application programming interface (API)~\cite{kalucki_streaming_2010}, five one-week sampled public tweets from September 2009 to July 2010 were selected. From the one-week samples composed of around 16.2 million to 57.6 million tweets representing about 15\% of public tweets~\cite{kalucki_streaming_2010}, nonzero-length messages were extracted which yielded about 52 million messages or utterances (see Ref.~\cite{alis_supplementary_2012} for datasets and parsing details). For better comparison with \textsc{pg} and \textsc{pgs} that have 50 subsets (authors) each, the weekly datasets were subdivided into ten groups of shuffled hourly data.

\textsc{subs} consists of about 14.7 million utterances from 15,809 movies provided by \texttt{opensubtitles.org}. The movie release years span from 1896 to 2010. See Ref.~\cite{alis_supplementary_2012} for parsing details and Ref.~\cite{alis_subs_movielist} for the complete list of movies.

\section{Utterance length distributions of datasets}
\begin{figure}[!ht]
\centerline{\includegraphics[scale=0.52]{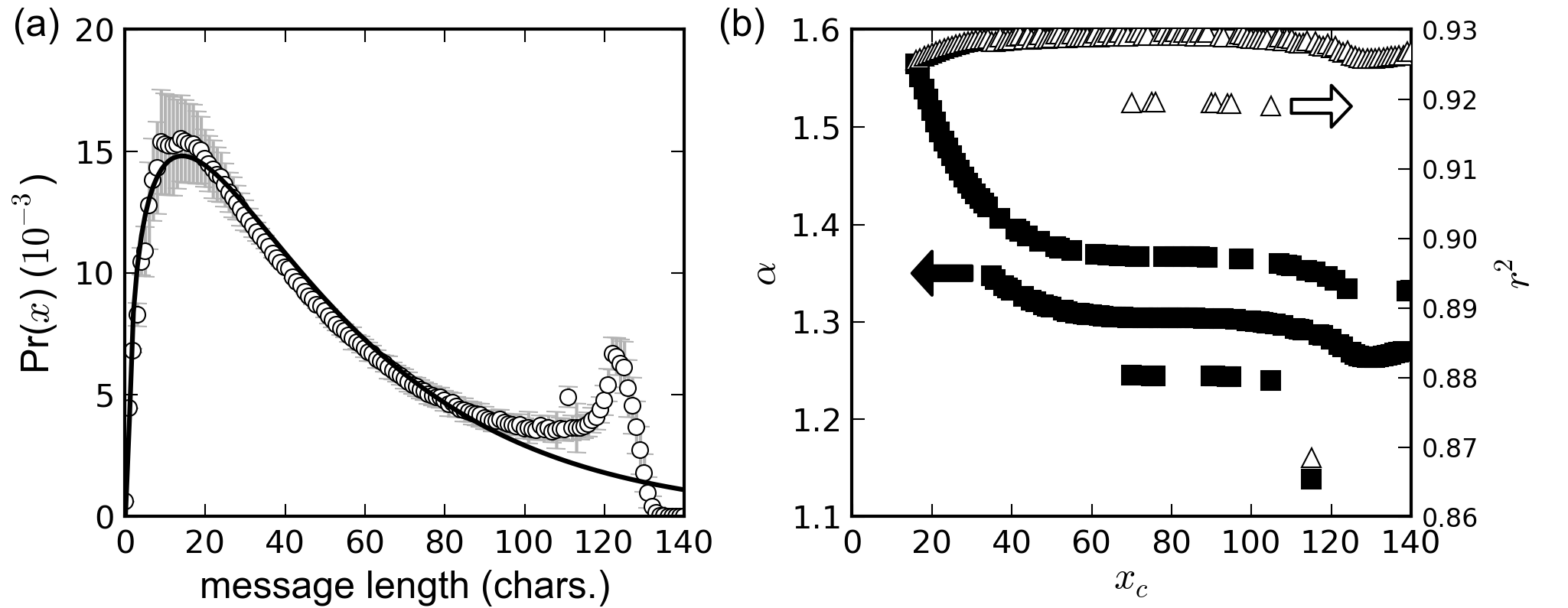}}
\caption{(a) Message length distribution of sampled tweets with the curve fit having the highest $r^2$ value  ($\alpha = 1.37$, solid line). Error bars are standard deviations from five one-week samples. (b) The $\alpha$ values (filled squares) of the fit from $x = 0$ to $x_c$ using Eq. (2) and its corresponding $r^2$ (unfilled triangles).
}
\label{fig:twitter-prob}
\end{figure}

Twitter conversations [Fig.~\ref{fig:twitter-prob}(a)] have an asymmetric and bimodal utterance length distribution. The left peak (mode) is at 16 characters which we take to be the natural distribution of message lengths i.e., it is the distribution of an unrestricted conversation. Similar to the argument used by Sigurd et al.~\cite{sigurd_word_2004} in their study of word and sentence length distributions of English, Swedish and German texts, and by Cancho and Sol\'{e}~\cite{cancho_least_2003} in their work on the origin of Zipf's law, we posit that the length of an utterance in a conversation is also governed by a trade-off between packing as much information as possible in an utterance and expressing the utterance as quickly as possible: the first objective is biased towards increasing length ($\sim x^{\alpha-1}$) while the other is biased towards decreasing it ($\sim e^{-x}$). Combining the two objectives, the following distribution is obtained: $\sim x^{\alpha - 1} e^{-x}$.

To account for a strict length limit for Twitter messages, the natural utterance length distribution was estimated by fitting a more general equation using a modified Levenberg-Marquardt least squares algorithm~\cite{jones_scipy:_2001} to the utterance length distribution from $x = 0$ to a cut-off length $x_c \in [16, 140]$ [Fig.~\ref{fig:twitter-prob}(b)],
\begin{equation}
\label{eq:gammafit}
\mathrm{Pr}(x) = \frac{\tilde{x}^{\alpha - 1} e^{-\tilde{x}}}{{\Gamma(\alpha)}},
\end{equation}
where ̃$\tilde{x} = (x-x_0)/s$ is the scaled utterance length $x$, while $\alpha$, $x_0$ and $s$ are fitting parameters that describe the shape, translation and ordinate scaling factor, respectively. This method of estimation assumes that the mixing parameter of the bimodal distribution is almost one in favor of the natural utterance length distribution. A bimodal distribution fitted using expectation maximization was not utilized because of a lack of an explicit model of the truncation distribution. Our goal is to estimate the median of the natural utterance length distribution so a resulting non-normalized unimodal distribution is acceptable.

\begin{figure}[!ht]
\centerline{\includegraphics[scale=0.52]{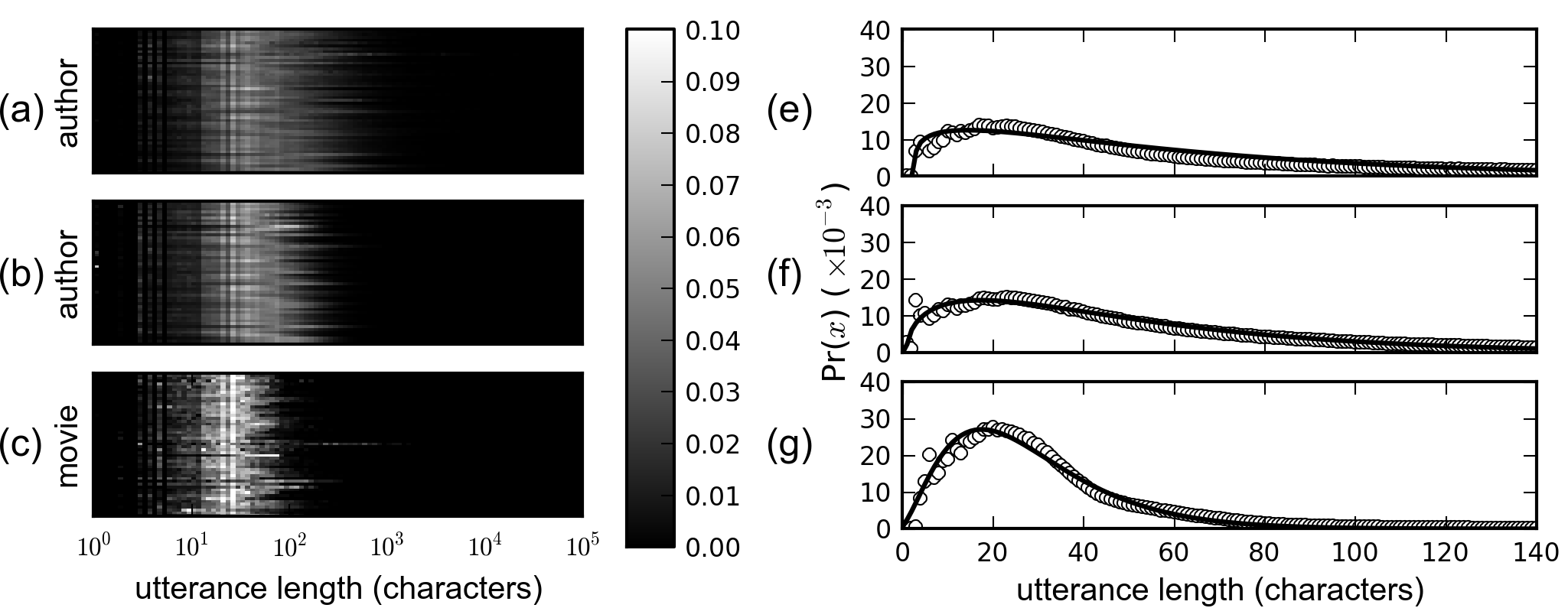}}
\caption{Utterance length distributions of (a) different authors in \textsc{pg} (b) different authors in \textsc{pgs} and (c) 50 randomly selected movies in \textsc{subs}. Distribution of utterance lengths over the entire (d) \textsc{pg}, (e) \textsc{pgs} and (f) \textsc{subs} datasets fitted with Eq. \eqref{eq:gammafit}.
 }
\label{fig:datasets-prob}
\end{figure}

When $\alpha$ approaches one, Eq.~\eqref{eq:gammafit} approaches an exponential distribution. The range of acceptable values of $\alpha \in [1.1, 1.6]$, [$r^2 \in (0.86, 0.93)$] for the Twitter dataset corresponds to a 57-order-of-magnitude increase in likelihood of finding an utterance length of $x = x_c = 140$ chars. compared to an exponentially decaying curve in the absence of a Twitter-imposed limit (see Ref. [27] for the fitting parameters distributions). However, another peak was found at 124 characters due to the 140-character limit, a limit that is absent in the other datasets, and is attributed to various tweet-shortening schemes. The absence of a length limit results to unimodal utterance length distributions for \textsc{pg}, \textsc{pgs} and \textsc{subs} [Fig.~\ref{fig:datasets-prob}].

Conversations in movies (interquartile range IQR = difference between the 3rd and 1st quartiles = 21 chars.) are of more uniform length than those in books (\textsc{pg} IQR median = 88 chars, \textsc{pgs} IQR median = 50 chars.). The much smaller \textsc{subs} IQR median compared to that of \textsc{twitter} (IQR median = 46 chars.) or that of its best fit of Eq. \eqref{eq:gammafit} (IQR median = 50 chars.) suggests that conversations in movies are less dependent on author style while the much larger IQR medians of \textsc{pg} and \textsc{pgs} point to a stronger dependence of these media on author style.

To minimize the effect of unequal author or movie utterances, and of noise due to differences in spelling and punctuation, Eq. \eqref{eq:gammafit} was fitted to \textsc{pg}, \textsc{pgs} and \textsc{subs} by computing for the normalized histogram of each author or movie then using the average probability for each utterance length as the probability density function to be fitted using least squares. Based on the fit of Eq. \eqref{eq:gammafit} ($\alpha = 1.48$, $x_0 = 0.862$, $s = 34.4$, $r = 0.984$), the \textsc{pgs} utterance length distribution [Fig. \ref{fig:datasets-prob}(e)] seems to be a horizontally compressed \textsc{twitter} best fit curve ($\alpha = 1.37$, $x_0 = 0.86$, $s = 36.4$) because of a smaller $s$ value. The \textsc{pg} utterance length distribution has a fatter tail [$1-F(140)=0.0896$; Fig. \ref{fig:datasets-prob}(d)] than that of the \textsc{pgs} utterance length distribution ($1 - F (140) = 0.0427$), and only its tail fits Eq. \eqref{eq:gammafit} quite well ($\alpha = 1.24$, $x_0 = 2.63$, $s = 48.6$, $r^2 =  0.970$). In contrast, the entire \textsc{subs} median distribution [$\alpha = 2.71$, $x_0 = −0.87$, $s = 10.7$, $r^2 = 0.988$; Fig. \ref{eq:gammafit}(f)] fits Eq. \eqref{eq:gammafit} and has almost no tail ($1 - F (140) = 1.19 \times 10^{−4}$). Thus, all datasets share the same distribution family as the Brown sentence length in words distribution further giving credence to the validity of the use of characters as a unit of utterance length.
 
\begin{figure}[!ht]
\centerline{\includegraphics[scale=0.52]{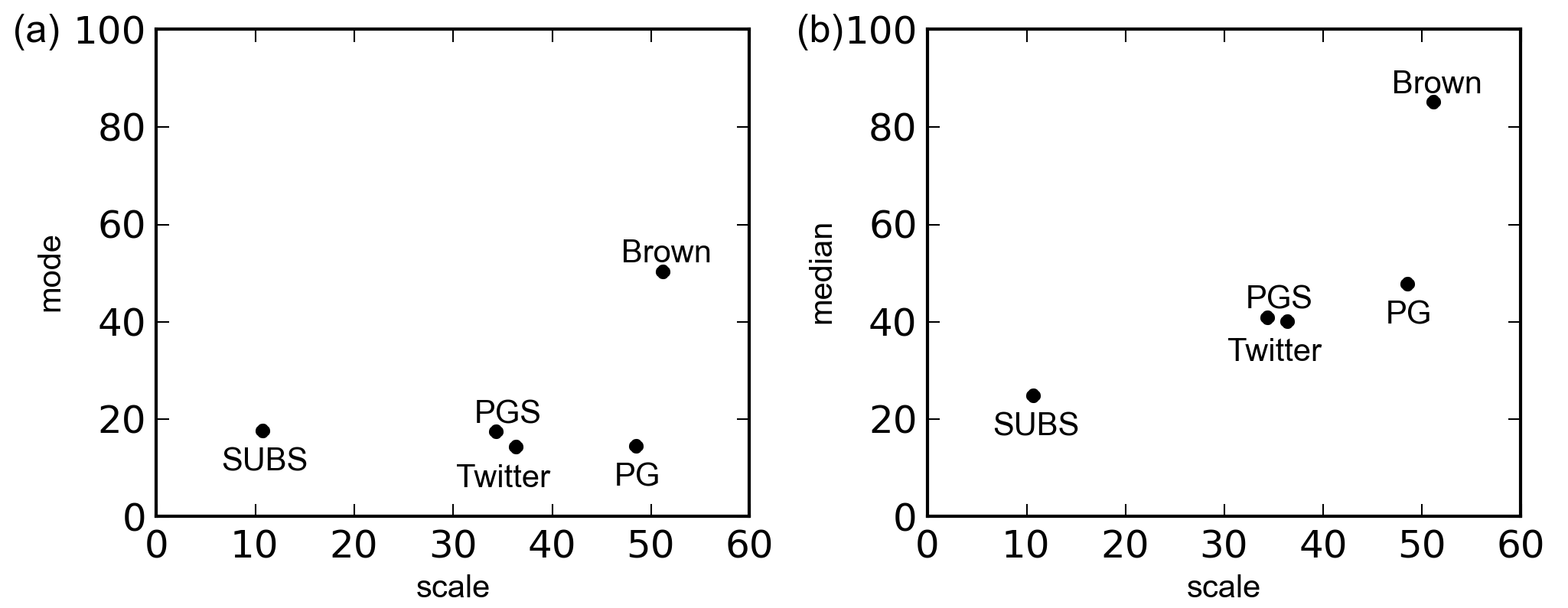}}
\caption{{\bf Mode and median of the distribution fits.} (a) Mode and (b) median of the fit of each distribution plotted against $s$.}
\label{fig:fit-desc}
\end{figure}

The mean length of utterance (MLU) is used to evaluate the level of language development of a child~\cite{klee_relation_1985,dollaghan_maternal_1999}. However, the use of the mean as a measure of central tendency is invalid because the utterance length distribution is very skewed to the right. The mode of a gamma distribution [Eq. \eqref{eq:gammafit}] is given by $(\alpha - 1)s + x_0$ but it does not appear to be correlated with $s$ [Fig. \ref{fig:fit-desc}(a)]. In contrast, the median, though not having a closed form equation for a gamma distribution, appears to be more correlated with $s$ [Fig. \ref{fig:fit-desc}(b)]: a larger median roughly implies a larger spread. The median, therefore, allows us to simultaneously describe both the location and scale of the utterance length distribution.

For the rest of this paper, the median utterance length and its median were used to describe each utterance length distribution. These measures are suitable for comparison between datasets because both are insensitive to outliers (robust) and do not assume a distribution (nonparametric). Any author dependence or deviation from a gamma distribution of the data would therefore not affect the results~\cite{alis_supplementary_2012}. Tests for significant differences were performed using the Mann-Whitney U test~\cite{mann_test_1947} with continuity correction because the distributions being compared are discrete and skewed.

\section{Utterance length and sample size}
\begin{figure}[!ht]
\centerline{\includegraphics[scale=0.7]{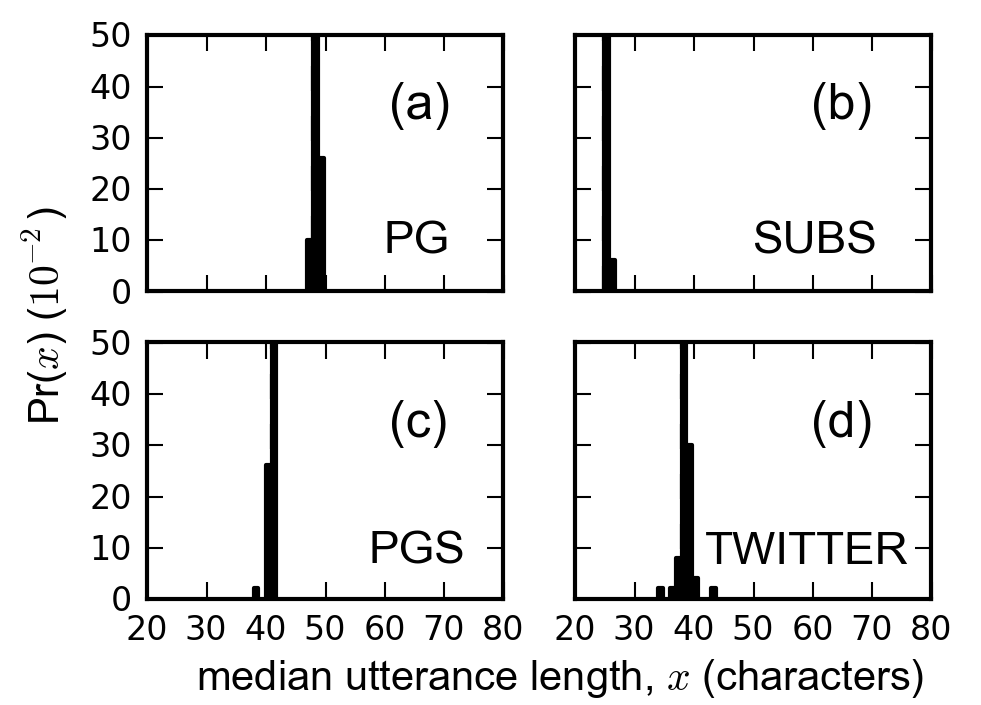}}
\caption{Distribution of median utterance lengths (median median utterance length: dashed lines) for (a) \textsc{pg}, (b) \textsc{subs}, (c) \textsc{pgs} and (d) \textsc{twitter}. The median utterance length data in (d) was estimated from the natural utterance length distribution of each \textsc{twitter} subset.
}
\label{fig:sampled-medians}
\end{figure}

\textsc{twitter}, \textsc{pgs} and \textsc{subs} were subsampled (with replacement) such that the sample size would be the same for each author's sample size in \textsc{pg}. By taking the distribution of subsample medians (Fig. \ref{fig:sampled-medians}) which is analogous to taking the distribution of sample means from normally-distributed data, we found that the median median utterance length (analogous to mean of sample means) of \textsc{subs} (25 chars.) is  very different from that of \textsc{twitter} (38 chars.), \textsc{pg} (48  chars.) and \textsc{pgs} (41 chars.).

Notably, the median median utterance length value of \textsc{subs} of 25 chars., which is not related to the existing maximum subtitle line length of 32-34 characters (Ofcom regulation~\cite{independent_television_commission_guidance_1999}), points to a fundamental difference in how the verbal medium is used in movies.

The median utterance length distribution of all datasets are significantly different from each other (see Ref. [27] for
 complete test results between each pair of dataset). Since the \textsc{pgs} median distribution is significantly different from the \textsc{subs} median distribution, conversational sentences in books are not the same as conversational sentences in movies though we posit that conversations in movies are closer to that of actual transcribed speech. \textsc{twitter} utterance lengths are stochastically smaller than \textsc{pg} and \textsc{pgs} but differ significantly from \textsc{subs} suggesting that Twitter is a less formal medium. We surmise that the smaller length is due to the more spontaneous and less formal tone of Twitter conversations than those in books.
 
\begin{figure}[!ht]
\centerline{\includegraphics[scale=0.7]{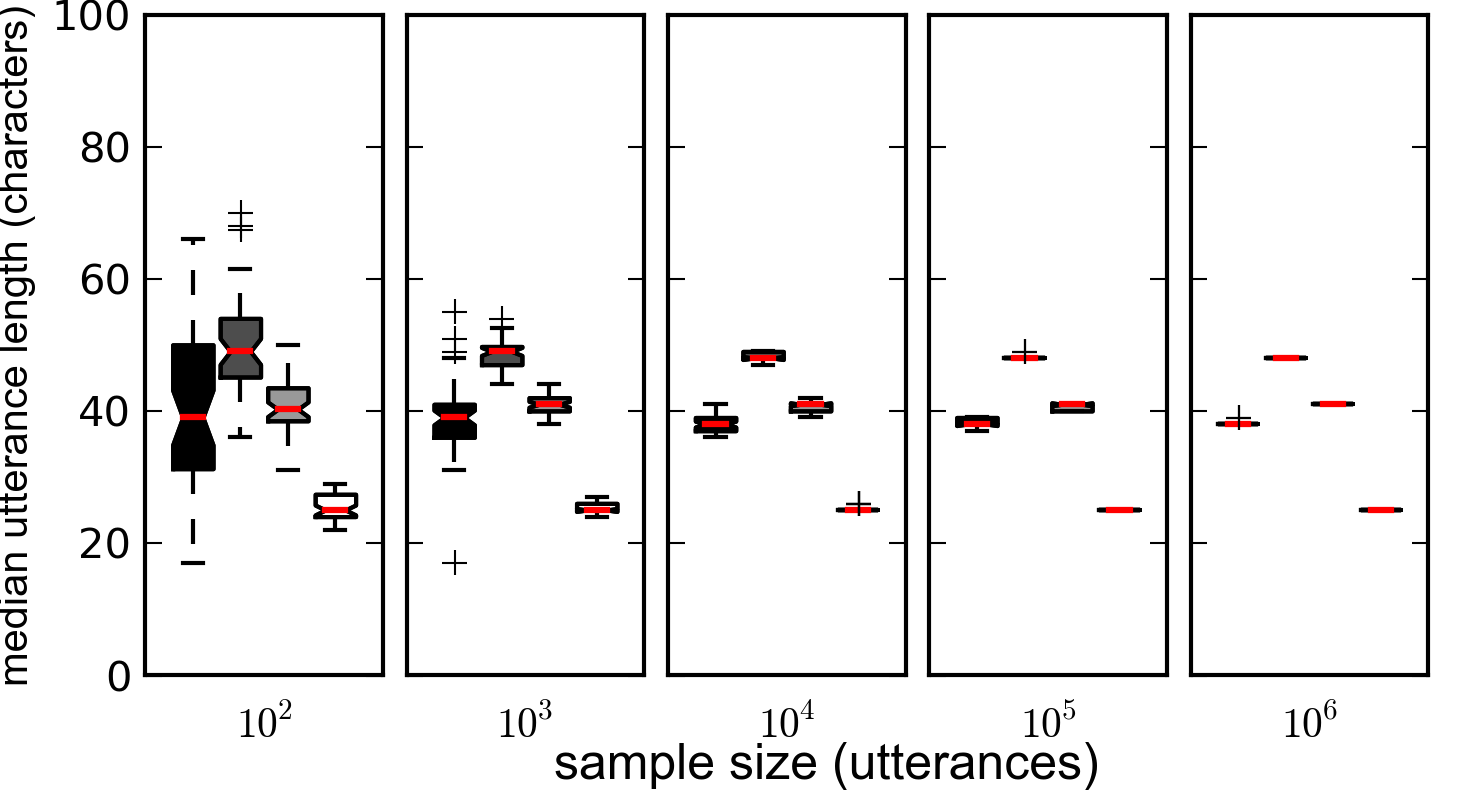}}
\caption{Distribution of median utterance length in subsampled \textsc{twitter} (black), \textsc{pg} (dark gray), \textsc{pgs} (light gray) and \textsc{subs} (unfilled).
}
\label{fig:dataset-medians}
\end{figure}

To investigate the effect of sample size $N$ on the median utterance length, each dataset was sampled (with replacement) into 50 groups each having $N$ utterances.  Similar to word frequency distributions that are dependent on $N$~\cite{bernhardsson_meta_2009}, the spread in, but not the location of, the medians distribution decreases as $N$ increases (Fig. \ref{fig:dataset-medians}) for all datasets. At $N = 10^5$ utterances, the median value of \textsc{subs} collapsed to a single value of 25 characters. At $N = 10^6$ utterances, \textsc{pg} and \textsc{pgs} collapsed to different
single median utterance length values of 48 and 41 characters, respectively, while \textsc{twitter} falls into two unique
 values of 38 and 39 characters.

The median utterance length distribution of \textsc{subs} is very different from the median utterance length distribution of the other datasets---it can be clearly distinguished from them even if the sample size is only $N = 100$ utterances (Fig. \ref{fig:dataset-medians}). \textsc{pg} and \textsc{pgs} median utterance length distributions are already distinguishable from each other but both overlap with \textsc{twitter} at $N = 100$ utterances. The \textsc{pg}, \textsc{pgs} and \textsc{twitter} median utterance length distributions do not overlap only at $N = 10^4$ utterances, thus giving us the required minimum sample size for meaningful comparison across communication media as a function of time (see Ref.~\cite{alis_supplementary_2012} for complete test results).
 
\section{Utterance length through time}
\begin{figure}[!ht]
\centerline{\includegraphics[scale=0.7]{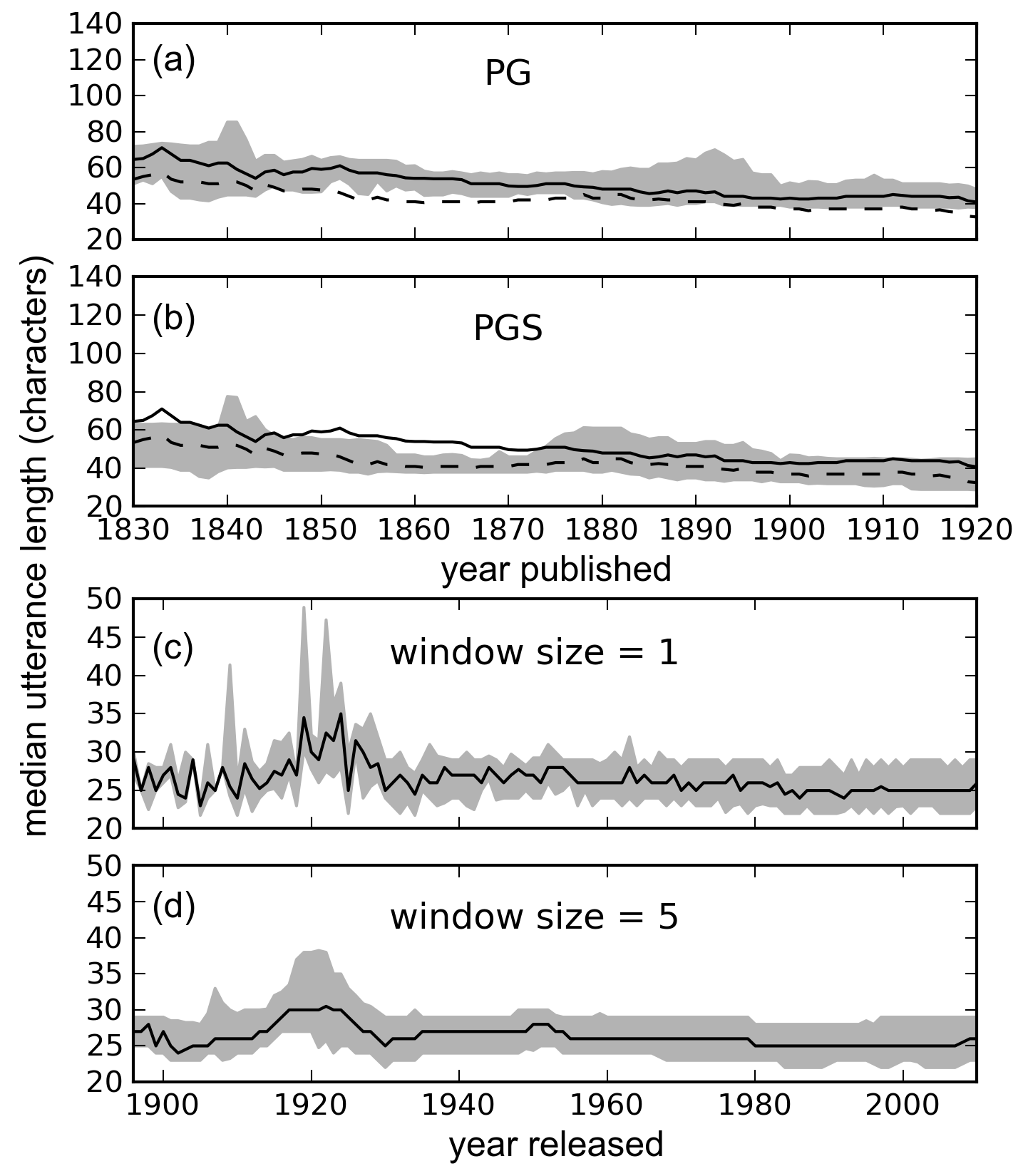}}
\caption{Median utterance length distribution of (a) \textsc{pg} and (b) \textsc{pgs} with window size of 10 years, and \textsc{subs} with window size of (c) 1 year and (d) 5 years. Only books with at least 1,000 utterances were considered. Publication years were retrieved from the US Library of Congress. The window sizes were selected so that the plots do not change appreciably when the window size is varied slightly. First to third quartiles (shaded), \textsc{pg} median
median utterance length (a-b, solid line), \textsc{pgs} median median utterance length (a-b, dashed line).
}
\label{fig:dataset-timeseries}
\end{figure}

The median median utterance length in both \textsc{pg} [Fig. \ref{fig:dataset-timeseries}(a)] (slope = -0.266 chars./yr, $r^2$ = 0.903, $p < 10^{-3}$ two-sided) and \textsc{pgs} [Fig. \ref{fig:dataset-timeseries}(b)] (slope = -0.189 chars./yr, $r^2$ = 0.814, $p < 10^{-3}$ two-sided) decreases with time but is not correlated with size (\textsc{pg} Spearman $\rho^2 < 10^{-3}$; \textsc{pgs} Spearman $\rho^2 = 0.00524$).

On the other hand, the median utterance length of \textsc{subs} [Fig. \ref{fig:dataset-timeseries}(c)] remains almost constant ($\sim27$ chars.) in time (slope = $-1.897 \times 10^{-3}$ chars./yr, $r^2$ = 0.121, $p < 10^{-3}$ two-sided) except for a conspicuous rise and increased spread in the median utterance length at around 1920 that does not flatten out even if the window size is increased from 1 year to 5 years [Fig. \ref{fig:dataset-timeseries}(d)]. The bump is likely due to the availability of ``talking pictures" and commercial television starting in the late 1920s. The silent movies prior to their release have a different ``conversation signature" from those of ``talkies".

The temporal behavior of \textsc{twitter} was not studied because \textsc{twitter} spans only a few weeks.

\section{Conclusion}
Though we do not usually notice the medium-dependence of conversations, we showed that conversations, as measured by orthographic utterance length, are slowly shortening in time within media but are drastically different across different media. These are fundamental differences that are effects not just of the milieu, but of the medium itself. Evolving technologies that lead to changes in communication media seemingly lead us to adapt our conversations, rather than such a technology suffering an early demise because it cannot adapt to our natural use of language. An extreme case in point is  the short message service (SMS) or ``texting." Originally designed with a character limit of 160 such that most sentences would fit in a single text message~\cite{milian_why_2009}, but with an ``access a letter via numerical keypad" constraint---it became a popular form of communication~\cite{_ictdata.org:_2010} with its own lingo~\cite{thurlow_daol:_2003}. Clearly, adaptation occurs with changing medium and sometimes with unexpected side-effects.

\begin{acknowledgement}
We thank the administrator of \texttt{opensubtitles.org} for providing us the text version of their English-languages movies subtitles. This work is supported by a grant from the UP Diliman-Office of the Vice Chancellor for Research and Development and by an Amazon AWS Education grant. 
\end{acknowledgement}

\end{document}